# Planar and van der Waals heterostructures for vertical tunnelling single electron transistors

Gwangwoo Kim[1], Sung-Soo Kim[2,*], Jonghyuk Jeon[2], Seong In Yoon[1], Seokmo Hong[3], Young Jin Cho[4], Abhishek Misra[5,6], Servet Ozdemir[5], Jun Yin[5], Davit Ghazaryan[5,7], Mathew Holwill[5], Artem Mishchenko[5], Daria V. Andreeva[8], Yong-Jin Kim[9], Hu Young Jeong[10], A-Rang Jang[1,3], Hyun-Jong Chung[4], Andre K. Geim[5], Kostya S. Novoselov[5], Byeong-Hyeok Sohn[2] and Hyeon Suk Shin[1,3,9, 11]

[1]Department of Energy Engineering, Ulsan National Institute of Science & Technology (UNIST), Ulsan 44919, Republic of Korea

[2]Department of Chemistry, Seoul National University, Seoul 08826, Republic of Korea

[3]Department of Chemistry, UNIST, Ulsan 44919, Republic of Korea

[4]Department of Physics, Konkuk University, Seoul 05029, Republic of Korea

[5]School of Physics and Astronomy, University of Manchester, Manchester M13 9PL, United Kingdom

[6]Department of Physics, Indian Institute of Technology Madras, Chennai, India

[7]Department of Physics, National Research University Higher School of Economics, Staraya Basmannaya 21/4, Moscow 105066, Russian Federation

[8]Department of Materials Science and Engineering, National University of Singapore, Singapore, 117575, Singapore

[9]Center for Multidimensional Carbon Materials, Institute of Basic Science (IBS), Ulsan 44919, Republic of Korea

[10]UNIST Central Research Facilities (UCRF), UNIST, Ulsan 44919, Republic of Korea

[11]Low Dimensional Carbon Material Center, UNIST, Ulsan 44919, Republic of Korea

*Present address: Carbon Composite Materials Research Center, Korea Institute of Science and Technology (KIST), Wanju 55324, Republic of Korea.

Correspondence and requests for materials should be addressed to H. S. S. (email: shin@unist.ac.kr ), B. H. S. (email: bhsohn@snu.ac.kr ) and K. S. N. (email: kostya@manchester.ac.uk ).

# Abstract

**Despite a rich choice of two-dimensional materials, which exists these days, heterostructures, both vertical (van der Waals) and in-plane, offer an unprecedented control over the properties and functionalities of the resulted structures. Thus, planar heterostructures allow p-n junctions between different two-dimensional semiconductors and graphene nanoribbons with well-defined edges; and vertical heterostructures resulted in the observation of superconductivity in purely carbon-based systems and realisation of vertical tunnelling transistors. Here we demonstrate simultaneous use of in-plane and van der Waals heterostructures to build vertical single electron tunnelling transistors. We grow graphene quantum dots inside the matrix of hexagonal boron nitride, which allows a dramatic reduction of the number of localised states along the perimeter of the quantum dots. The use of hexagonal boron nitride tunnel barriers as contacts to the graphene quantum dots make our transistors reproducible and not dependent on the localised states, opening even larger flexibility when designing future devices.**

# Introduction

Graphene quantum dots (GQD) and graphene nanowires have been attracting attention because of the linear spectrum obeyed by the quasiparticles[1,2] (zero mass allows one to reach large quantisation energy, comparable with the room temperature, for relatively large quantum dots[3,4]), small spin-orbit interaction[5,6], good chemical stability[7] and the ability to support high currents. At the same time, the transport properties of such quantum dots, which are typically carved out of large sheets of graphene, are dominated by the localised edge states[1,8,9]. Furthermore, arranging tunnelling contacts (usually carved graphene constriction) to such quantum dots with the specific, reproducible conductivity is a separate challenge.

Here we use planar[10-12] and vertical[13-15] heterostructures to mitigate the issues with the localised states both at the edges of the quantum dots and at the edges of the contacts. We propose to form GQDs inside the hexagonal boron nitride (hBN) matrix through catalytic-assisted substitution of boron and nitrogen atoms by carbon[16]. Since the lattices of graphene and hBN are very similar – most of the bonds in a GQD become properly terminated which minimizes the number of such localised states. Thus, the edges of a GQD become effectively passivated with hBN. We would like to stress that our method allows the formation of GQDs with specific size, and can be extended to other structures and devices. We then used hBN tunnelling barrier and graphene electrodes in order to form contacts to such quantum dots, creating single electron tunnelling transistors. Such method results in very precise, reproducible contact resistance, and the graphene-hBN interface free of localised states.

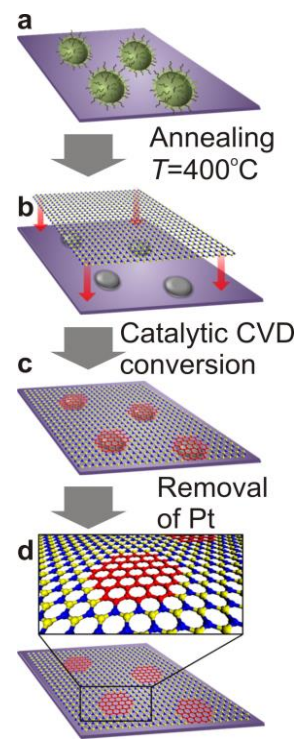

**Figure 1|** The fabrication steps of GQD-hBN in-plane heterostructure based on hBN to graphene conversion catalysed by Pt NPs. **a,** The self-assembly of diblock copolymer micelles PS-P4VP with $H_2PtCl_6$ on $Si/SiO_2$ substrate. **b,** Transfer of hBN monolayer on $SiO_2$ substrate covered by Pt NPs (blue spheres – boron atoms, yellow spheres - nitrogen). **c,** Formation of the GQDs on top of an array of Pt NPs by catalytically-assisted CVD (red spheres – carbon atoms). **d,** The obtained in-plane GQD-hBN heterostructure after the removal of Pt NPs.

## Results

### Formation of graphene quantum dots

In order to fabricate the in-plane graphene/hBN heterostructures, we used a conversion reaction on a patterned Pt-SiO$_2$ substrate described in[16]. Based on the spatially controlled conversion, the growth of in-plane GQD-hBN heterostructure was achieved on a SiO$_2$ substrate covered by an array

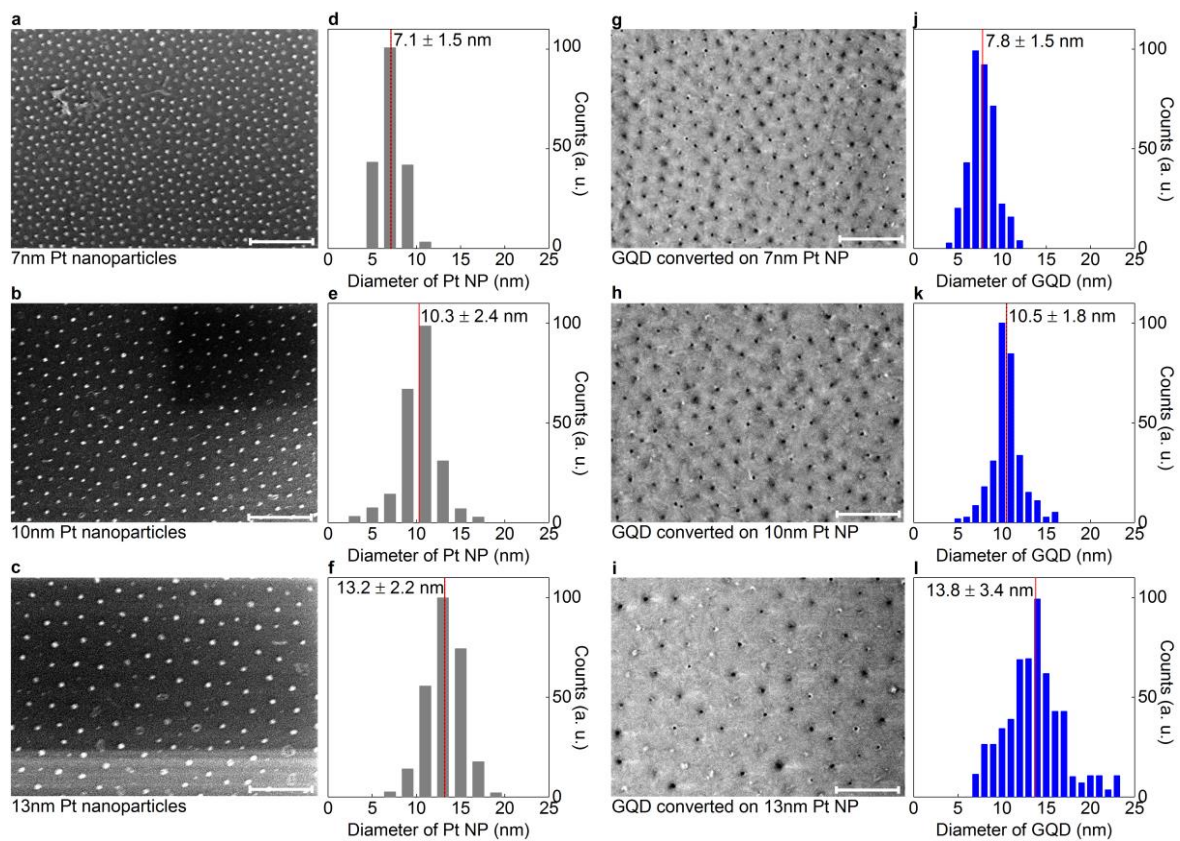

**Figure 2| Size and spatial distribution of the GQDs. a-c,** The SEM images of 7, 10, and 13 nm sized self-assembled arrays of Pt NPs on SiO$_2$ substrates, respectively. Scale bar 300nm. **d-f,** Corresponding size distribution histograms of Pt NPs on SiO$_2$ substrates. Numbers give the average (marked by red lines) and the standard deviation. **g-i,** The SEM images of GQD-hBN in-plane heterostructures prepared on pristine SiO$_2$. Scale bar 300nm. **j-l,** Corresponding size histograms of GQD-hBN samples. Numbers give the average (marked by red lines) and the standard deviation.

of platinum (Pt) nanoparticles (NP), as illustrated in Figure 1. The high-quality hBN monolayer was first grown on Pt foils via chemical vapour deposition (CVD), using ammonia borane as a precursor[17] (the experimental details for the growth and the characterisation of monolayer hBN are provided in Methods and Supplementary Figure 1). Next, the hBN film was transferred onto an array of Pt NPs spread over SiO$_2$ substrate, prepared with the aid of self-patterning diblock copolymer micelles[18]. The Pt NPs were obtained by spin-coating of a single-layer of polystyrene-block-poly(4-vinyl pyridine) (PS-P4VP) micelles, with an H$_2$PtCl$_6$ precursor for Pt NPs within their cores, followed by the annealing at 400°C. Then, the conversion of the hBN sheet to graphene on the array of Pt NPs on the SiO$_2$ substrate was accomplished at ~950°C in methane/argon atmosphere. During the reaction, the hBN on top of Pt NPs was selectively converted to graphene, with the formation of uniform GQD arrays embedded in the hBN film (Supplementary Figure 2). Notably, depending on the molecular weight of the diblock copolymer, the size of the Pt NPs was controlled in the range of 7 to 13 nm. The scanning electron microscopy (SEM) images presented in Figure 2a-c demonstrates the uniform arrays of Pt NPs with diameters of approximately 7, 10, and 13 nm (Figure 2d-f). Next, the as-prepared GQD-hBN in-plane heterostructure was placed in aqua regia solution to remove Pt NPs (Figure 2g-I), and finally, transferred onto arbitrary substrates for further characterisation and processing. Note, that the area of obtained GQDs is comparable to the size of the Pt NPs, as shown in Figure 2. The removal of the Pt NPs was confirmed by XPS and TEM (Supplementary Figures 3-4)

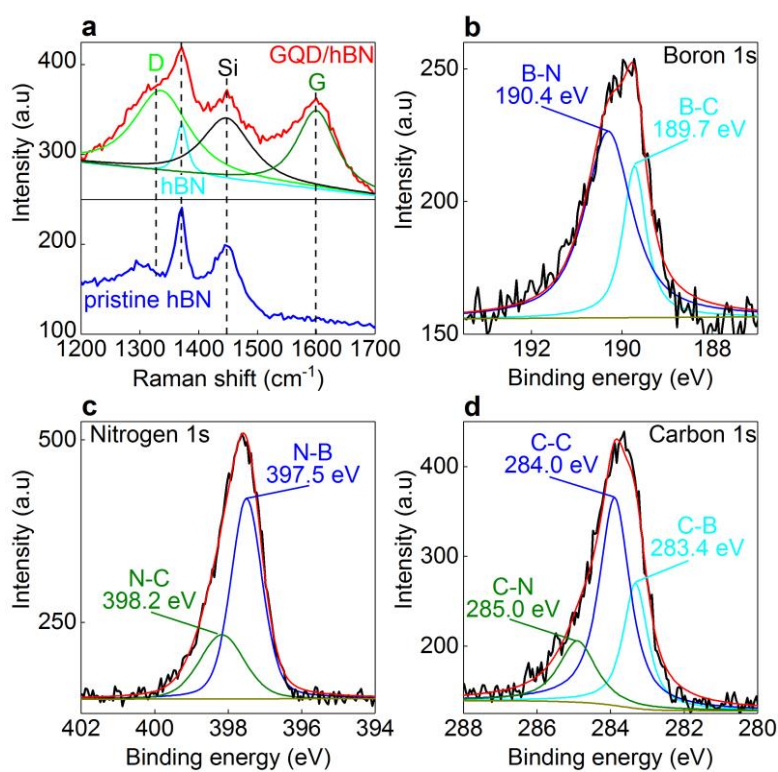

**Figure 3| Characterisation of GQD-hBN interface of the in-plane heterostructure with GQDs of the size of 7nm. a,** Raman spectra of GQD-hBN planar heterostructure (red) and pristine hBN (blue). XPS spectra of GQD-hBN planar heterostructure: **b,** boron 1s, **c,** nitrogen 1s, and **d,** carbon 1s spectrum.

The formation of GQDs was confirmed by Raman spectroscopy (Figure 3a) and EELS (Supplementary Figure 5). Typical Raman signals of GQDs and hBN were observed from the in-plane heterostructure of GQD-hBN transferred onto a SiO$_2$ substrate: the D (1345 cm$^{-1}$) and G (1595 cm$^{-1}$) bands of graphene with an intervening E$_{2g}$ peak (1371 cm$^{-1}$) of an hBN[19,20]. Furthermore, the in-plane graphene domain size ($L_a$) was calculated using the ratio of the integrated intensity ($I_D/I_G$) according to the Tuinstra-Koenig relation[21] (Supplementary Table 1 and Supplementary Note 1), which is consistent with the size of GQDs, observed by SEM, Figure 2. The formation of GQD-hBN heterostructures was also confirmed from absorption bands of hBN and GQDs in the UV-vis absorption spectra (Supplementary Figure 6) and EELS mapping (Supplementary Figure 5).

In order to characterise the interface between the GQDs and hBN in our in-plane heterostructures, we performed an X-ray photoelectron spectroscopy (XPS), Figure 3b–d, which suggests a reasonable formation of bonds among carbon, nitrogen, and boron atoms[22,23]. The obtained boron 1s peak, illustrated in Figure 3b can be deconvoluted into two peaks with the energies of 189.7 eV and 190.4 eV, which are attributed to the B-C and B-N bonds, respectively. Notably, such a peak value of the B-N bond is very close to the measured value of boron 1s (190.5 eV) in the pristine hBN monolayers (see Supplementary Figure 1). Next, the nitrogen 1s peak, presented in Figure 3c is composed of two peaks, corresponding to N-B bonds (397.5 eV) and N-C bonds (398.2 eV). Finally, the formation of C-B (283.2 eV) and C-N (285.0 eV) bonds were confirmed in the XPS carbon 1s spectrum (Figure 3d). In addition, we also confirmed the C-N bonds at 1273 $cm^{-1}$ and the B-N bonds at 1375 $cm^{-1}$ in the measured infrared (IR) spectra of the GQD-hBN in-plane heterostructure (Supplementary Figure 7). The B-C bonds, which are expected to appear at approximately 1020 $cm^{-1}$, were not identified due to an emergence of a very strong $SiO_2$ peak. Due to the small size of our quantum dots - the lattice mismatch between graphene and hBN is not expected to lead to the formation of dislocations as in the case of bulk graphene/hBN planar heterostructures[24].

**Vertical single electron tunnelling transistors**

Such GQDs embedded in an hBN matrix are ideally suited for the formation of van der Waals heterostructures[25-27]. To this end, we prepared vertical tunnelling single electron transistors[28-31], where the transparency of the contacts is controlled by tunnelling through atomically thin hBN layers, thus avoiding the issue of the localised states in the contacts[8]. Our van der Waals heterostructures of the type 30nm_hBN/Gr/2hBN/GQD-hBN/2hBN/Gr/20nm_hBN have been assembled on Si/$SiO_2$ substrate (acting as a back gate) by using dry transfer method[32] (see device fabrication in Methods). Here, 30nm_hBN stands for the hBN layer with an approximate thickness of 30nm, Gr – for graphene, 2hBN – for 2-layer thick exfoliated hBN, GQD-hBN – for GQD-hBN lateral heterostructure. Schematic structure and the layer arrangement of our devices is presented on the inset to Figure 4e. The GQD-hBN layer was sandwiched between two thin hBN layers to isolate the quantum dots from the contacts to ensure a long lifetime of electrons within the quantum dots, thus, to allow the detection of the single electron energy levels. All our devices were fabricated in a symmetric configuration - with the same number of hBN layers on each side of the GQD-hBN layer. Devices with two (2hBN) and three (3hBN) layer thick hBN tunnelling barriers have been produced. Information about all the devices measured can be found in the Supplementary Table 2.

We performed tunnelling spectroscopy on our van der Waals heterostructures, by applying a mixed signal of AC and DC bias voltages between the two graphene electrodes, and a gate voltage to the silicon substrate (see Methods and[33] for details). Figure 4 presents a colour map of differential conductance $G(V_g, V_b)=dI/dV_b$ as a function of the gate ($V_g$) and bias ($V_b$) voltages for one of our devices with 13nm GQDs (examples of the tunnelling conductance for devices with 7nm GQD are presented in Supplementary Figure 8). Two types of sharp peaks can be identified on top of the smooth background; (i) those organised into overlapping diamonds (the edges of the diamonds are marked by thin long red arrows, Figure 4a), (ii) those following square root dependence (marked by thin long green arrows, Figure 4a-c). We attribute the peaks with square root dependence in $G(V_g, V_b)$ to arise due to tunnelling through the localised electronic state, and the diamond-shaped features - to Coulomb blockade diamonds owing to tunnelling through single electron states in individual graphene quantum dots.

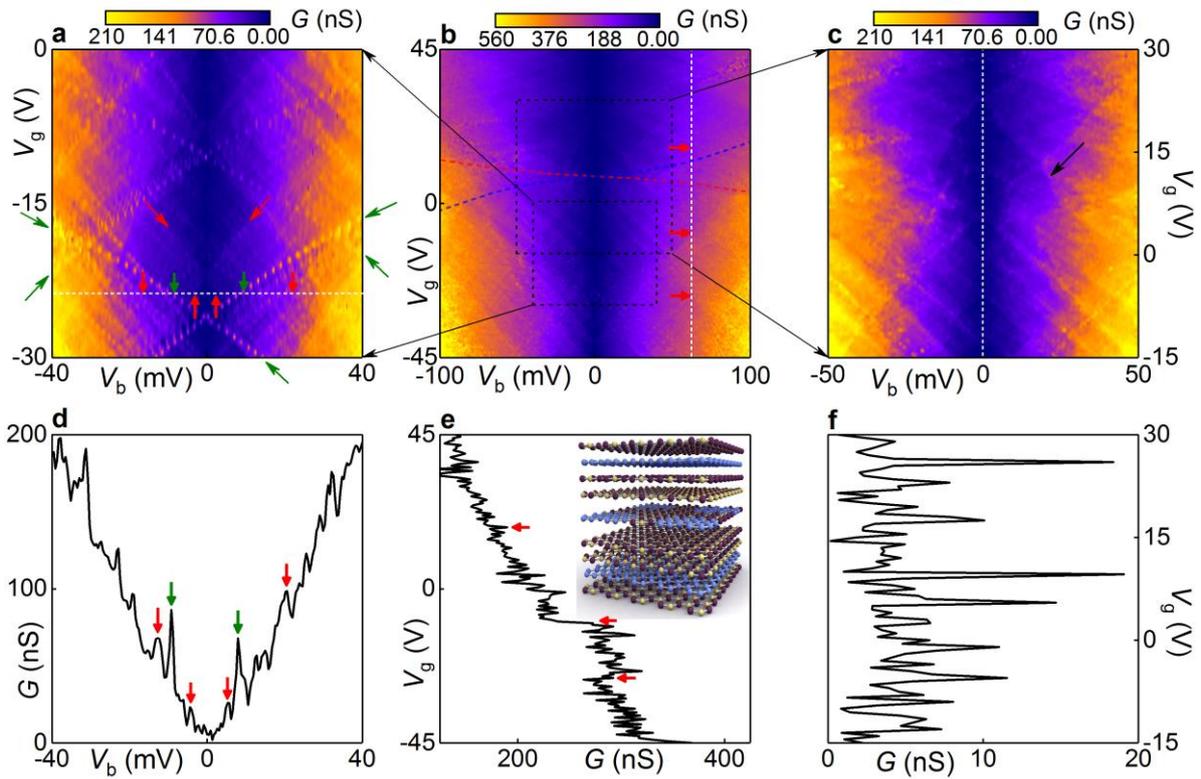

**Figure 4| Multi-channel single electron tunnelling transistors based on the GQDs of 13nm in size. a,** The low excitation measurements of low bias region of (b). Thin and long green arrows indicate the tunnelling events through the localised states in the middle hBN layer. Thick and short arrows indicate the edges of the Coulomb diamonds (red), and the resonances from localised impurity states (green) crossing the white dashed line ($G(V_g$=-24V, $V_b$), presented on panel (d)). **b,** Conductance $G(V_g,V_b)$ measured at $T$=250mK. The red (blue) dashed line mark the event of the Fermi level in the top (bottom) graphene layer aligning with the Dirac point. Red arrows indicate the edges of the Coulomb diamonds crossing the white dashed line ($G(V_g, V_b$=63mV), presented on panel (e)) **c,** The magnified plot of (b) denoting the peculiar shape of Coulomb diamonds when the Fermi level in one of the graphene contacts aligns with the Dirac point (black arrow). The dashed white line indicates the cross-section presented in (f). **d,** The plot of conductance $G(V_g$=-24V,$V_b$) from panel (a). Thick and short arrows indicate the same events as in (a): the edges of the Coulomb diamonds (red), and the resonances from localised impurity states (green). **e,** The conductance $G(V_g, V_b$=63mV) plot. Red arrows indicate the same events as in (b): edges of the Coulomb diamonds. The inset shows the schematic structure of the van der Waals stack. Graphene contacts are separated by hexagonal boron nitride layers from the GQD-hBN layer. Carbon atoms are blue, boron – yellow, nitrogen – purple. **f,** The conductance $G(V_g, V_b$=0mV) plot from panel (c) (marked by white dashed line).

Each diamond corresponds to a Coulomb blockade regime in one particular GQD. For a single GQD, one would observe a sequence of diamonds which connect to each other only at vertices. Since we have a large number of GQDs connected in parallel – we observe a number of overlapping diamonds[34]. The zero-bias conductance within the diamond, Figure 4f is given by the background tunnelling through the 5 layer hBN (2 layers of hBN on each side of the middle GQD-hBN layer) and is within the expected range[35], indicating the absence of the pin-holes in the barrier. To confirm that Coulomb diamonds are originating from the single electron charging events happening at GQDs, heterostructures of the stack 30nm_hBN/Gr/2hBN/CVD-hBN/2hBN/Gr/20nm_hBN were produced. In such devices, the same CVD grown hBN as used in devices of Figure 4 is utilised, however, this CVD hBN does not contain any GQD in it. No Coulomb diamonds were observed in such devices (see Supplementary Figure 9). Furthermore, the conductivity of such devices is significantly lower than that for devices with GQDs, which proves that the Coulomb diamonds we observe are indeed coming from the GQDs.

In our diagrams (Figure 4), each set of diamonds corresponds to a particular graphene quantum dot. Thus, we can estimate the number of GQDs involved in tunnelling, which gives us approximately 40 quantum dots connected in parallel for the device with 13nm GQD. Thus, the number of GQDs we see participating in tunnelling is much smaller than the total number of GQDs within the area of the device (~50 GQDs per $\mu m^2$, the total area of the device ~30 $\mu m^2$ for the sample with 13nm GQD). Currently we don't have an explanation for this effect. However, it can be speculated that as it is the silicon gate (separated from the layer with GQD by approximate 300nm of $SiO_2$ and hBN), which provides the most efficient screening (graphene electrodes provide only weak screening) – GQD interact strongly between themselves via Coulomb interaction. It means that the observed Coulomb diamonds are the result of the collective behaviour of several GQDs within the 300nm radius. This would also explain the different intensities of the conductivity peaks, Figure 4d,f.

**Low-density graphene quantum dots**

In order to avoid the large number of GQDs to be connected in parallel, thus obscuring the Coulomb diamonds, hBN layer with a low density of graphene islands was prepared. To this end we used a strongly diluted micellar solution to achieve a low concentration of $H_2PtCl_6$: 1 mL of the PS-P4VP copolymer solution with $H_2PtCl_6$ was diluted by 400 mL of pure PS-P4VP. Such mixed solution was spin-coated on the $SiO_2$ substrate, and the micellar film was treated by oxygen plasma to produce Pt NPs. The hBN monolayer was transferred onto the Pt NPs/$SiO_2$ substrate, and the conversion reaction was performed for conversion of hBN on Pt NPs to graphene. After the aqua regia treatment to remove Pt NPs, a GQD-hBN monolayer with a relatively long spacing (0.5 to 1.5 μm) between GQDs (Figure 5a) was obtained. Note, that this method gives a non-uniform distribution of GQDs.

We used such hBN with low density of GQDs to prepare single electron tunnelling transistors 30nm_hBN/Gr/2hBN/GQD-hBN/2hBN/Gr/20nm_hBN as it has been described above (see inset to Figure 4e). The conductance of one of such devices (with the active area of 30μm$^2$) as a function of the gate and bias voltages is presented in Figure 5b-d. Note, that the characteristic conductance for such a device is at least an order of magnitude smaller than that for devices with periodic, high-density arrays of GQDs (see the data presented on Figure 4, note that the areas and the barrier thickness for these devices are the same). This is because the tunnelling now occurs through a smaller number of GQDs. At the same time, the Coulomb diamonds are visible much clearer in such devices (Figure 5d) partly because of smaller number of overlapping Coulomb diamonds due to smaller number of GQD, and partly because each GQD now act independently, interacting only weakly with other GQDs, since the distance between them is larger than the distance to the gate.

# Discussion

The schematics of the formation of the diamonds are presented in Figure 6a-c and Supplementary Figures 11 and 12. When the size quantisation levels in the GQD are positioned outside of the bias window – no current flows through the quantum dots. At positive biases, a finite conductance is observed once the size quantisation level is below the Fermi level in the top graphene (such events are modelled by red lines in Figure 6d) and above the Fermi level in the bottom graphene (modelled by blue lines in Figure 6d). The combination of four of such lines gives a diamond of low conductivity. If the Fermi level in one of the graphene contacts is close to the Dirac point, where the density of states vanishes – the electrostatics dictates[30] that the edges of the diamonds will not be straight lines anymore and will have square root dependence in the $V_g$-$V_b$ coordinates (see Figure 5d and the Supplementary Note 2 for the details of the model). Such events are indeed observed in our measurements (marked by the black arrow in Figure 4c, also see Supplementary Note 2 and Supplementary Figure 10).

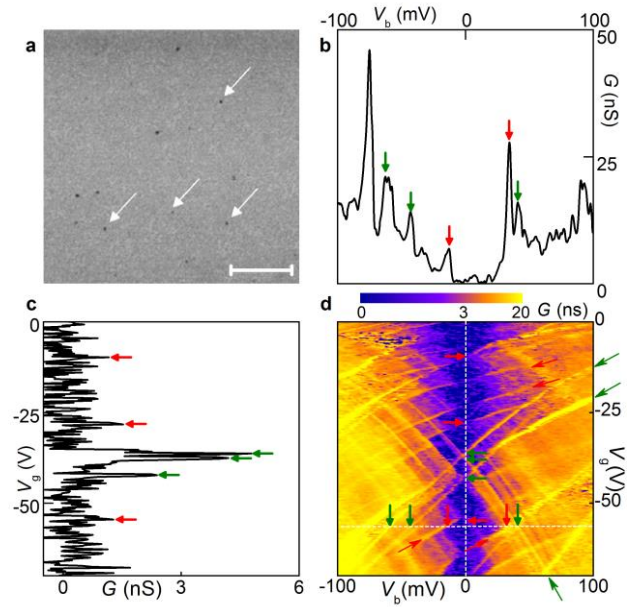

**Figure 5| Low density non-periodic array of 10 nm GQDs embedded in hBN matrix and electron tunnelling transistors based on such GQDs. a**, SEM image of a GQD-hBN sample obtained after the transfer of hBN monolayer on Pt NPs/SiO$_2$ substrate and the conversion reaction. It shows GQDs with a long spacing (0.5 to 1.5 μm), marked by white arrows. Scale bar 1μm. **b**, The conductance $G(V_g$=-56V,$V_b)$ plot (extracted from (d) along the horizontal white dashed line). Arrows indicate peaks originating from boundaries of Coulomb diamonds (red) and from impurity assisted-tunnelling (green). **c,** The conductance $G(V_g$, $V_b$ =0mV) plot (extracted from (d) along the vertical white dashed line). Arrows indicate peaks originating from boundaries of Coulomb diamonds (red) and from impurity-assisted tunnelling (green). **d**, Conductance $G(V_g,V_b)$ for a device with aperiodic 10nm GQDs measured at $T$=250mK. Thick and short arrows indicate the crossing of the edges of the Coulomb diamonds (red arrows) and the impurity–assisted tunnelling peaks (green) with the white dashed lines. The positions of the arrows are the same as on panels (b), (c). Thin and long arrows indicate the edges of the Coulomb diamonds (red arrows) and the impurity-assisted tunnelling peaks (green arrows).

The width of the diamonds in $V_b$ gives a characteristic charging energy required to place an extra electron into a quantum dot. Experimentally, the easiest way to determine the width of the diamonds is by taking constant bias cross-sections (such as presented in Figure 4e) of the conductivity plot $G(V_g, V_b)$, where the edges of the diamonds are seen as distinct steps (marked by red arrows on Figure 4b,e). The bias at which such steps disappear is then taken as the width of the diamonds. For our 13nm (Figure 4), 10nm (Figure 5) and 7nm (Supplementary Figure 8a,b) quantum dots the charging energy was found to be of the order of 80±15meV, 100±15meV and 160±20meV respectively, well in line with what is expected for the size quantisation for quantum dots of such a diameter[36-38]. The fact that the size quantisation energy scales as expected with the size of the GQD serves as an additional argument that the tunnelling occurs through the states in the quantum dots.

Simultaneously with the characteristic diamonds, a few conductance peaks which approximately follow the square root behaviour have been observed. The square root behaviour is coming from the linear density of states in the graphene electrodes and the fact that, due to the small density of states the bottom graphene electrode doesn't entirely screen the electric field from the gate[39]. We attribute these features to the tunnelling through localised states in the central GQD layer[40,41]. Each localised state produces two lines in $G(V_g, V_b)$ – when it aligns with the Fermi level in the bottom and in the top graphene contacts. The energy positions of these lines can be fitted with very high precision (Figure 6d). From such a fitting we can extract the energy position of the localised states with respect to the Dirac points in the graphene layers as well as their spatial positions in the barrier. Thus, we found that energetically all the localised states observed are situated in the range of ±150meV in the vicinity of the Dirac points in the contacts. Our fitting also confirms that spatially all the localised states are indeed located in the central layer[42] (hBN with GQDs). Note that it has been demonstrated that impurity states in the middle of the barrier contribute the strongest to the tunnelling current[43] (see Supplementary Figure 13 for examples of phonon- and impurity-assisted tunnelling). The number of localised states we can see in our devices is very low (between 3 and 6, depending on the particular device, see Supplementary Table 2) much lower than the number of the GQD observed. This suggests that the edges of our GQDs are well passivated and do not produce additional localised states.

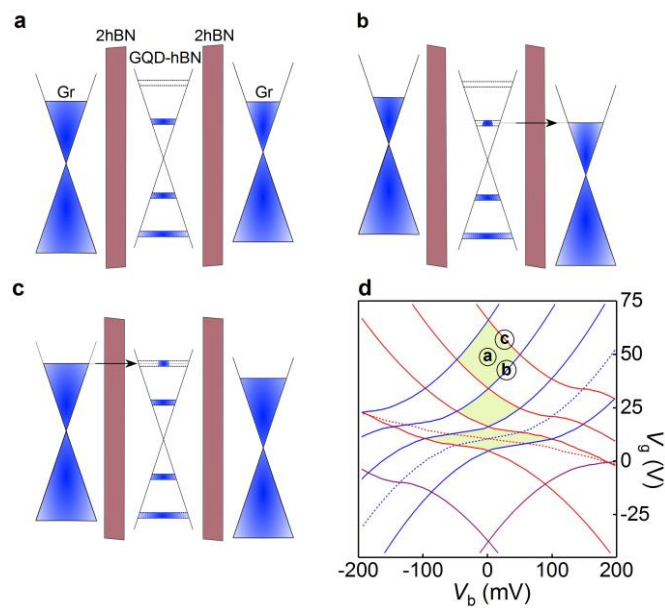

**Figure 6| Modelling of the single electron charging effect. a-c,** Schematic representation of a single electron charging effect. The corresponding electrostatic lines are denoted in (d). **d,** Modelling example of the alignment of the different energy levels in a device with 13nm GQD (see Methods and Supplementary Notes 2 for the details of the modelling). Red (blue) dashed lines - Fermi level in top (bottom) graphene electrode aligning with the Dirac point. Purple lines – Fermi levels in the graphene contacts being aligned with the localised state located in the middle hBN layer with energy 140 meV below the Dirac point. The set of solid red (blue) lines correspond to single electron energy levels in GQD aligning with the Fermi level in the top (bottom) electrode. Space between four of such lines forms a Coulomb blockade diamond. Note the distorted shape of the diamond when the Fermi level in the contacts passes through the Dirac points.

In conclusion, we demonstrated a way of synthesis of GQDs embedded in the hBN matrix. Such GQDs exhibit a very low number of edge states. The geometry allows easy incorporation into van der Waals heterostructures, where we demonstrate single electron tunnelling transistors. Our approach

– the combination between the in-plane and van der Waals heterostructures – allows the fabrication of high quality graphene quantum dots for transport experiments. The in-plane heterostructures allow fabrication of graphene quantum dots without the dangling bonds and localised states at the perimeter. At the same time, the van der Waals heterostructures allow fabrication of controlled tunnelling barriers, again without any localised states. We hope that our approach will pave the way for many other types of devices and physical phenomena to be studied.

## Methods

**The growth of GQDs embedded in the hBN sheet**

The single layer of hBN was synthesised on Pt foil using ammonia borane as a precursor by the CVD method. Experimental details on the synthesis of CVD-grown hBN on Pt can be found in a previous report[17]. The Pt NPs array on a $SiO_2$ substrate was prepared using self-patterning diblock copolymer micelles[18]. A single-layer of polystyrene-block-poly (4-vinylpyridine) (PS-P4VP) micelles with $H_2PtCl_6$, a precursor for Pt NPs, in their cores was spin-coated on the $SiO_2$ substrate. To fabricate the Pt NP array, the micellar film on the $SiO_2$ was annealed at 400°C for 30 min in air. The hBN layer was transferred onto the Pt NPs/$SiO_2$ substrate using a wet-transfer method (electrochemical delamination). Then, the hBN layer transferred on Pt NPs/$SiO_2$ was loaded into the centre of a vacuum quartz tube in a furnace for the conversion reaction[16]. The tube was pumped down to 0.21 Torr with pure argon gas (50 sccm). Then the furnace was heated to 950 °C in 40 min. During the reaction, methane gas (5 sccm) with argon (50 sccm) was flown as the source for graphene growth. During the reaction, the hBN region on the Pt NPs was converted to graphene, and after 10 min of growth, a uniform GQD array embedded in the hBN film was obtained.

**Transfer method**

The GQD-hBN film on Pt NPs/$SiO_2$ could be transferred to any other substrate via wet-transfer method using HF and an aqua regia solution. First, polystyrene (PS) was spin-coated on the sample, and it was immersed in an HF solution (5% in DI water) to detach the GQD-hBN film. The floating PS film was then transferred to the aqua regia solution (3:1 mixture of hydrochloric acid and nitric acid) to remove Pt NPs and then rinsed with copious amounts of deionised water (DI). Finally, the film was transferred onto the target substrate and PS was removed with toluene to obtain a GQD-hBN film on the substrate.

**Characterisation**

Scanning electron microscopy (Verios 460, FEI) and atomic force microscopy (Dimension Icon, Bruker) were used to determine the surface morphology of the samples. Raman spectra were measured using a micro Raman spectroscope (alpha 300, WITec GmbH) using 532 nm. The UV-vis absorption spectra of the GQD-hBN samples were recorded on a Cary 500 UV-vis-near IR spectroscope, Agilent. X-ray photoelectron spectroscopy (K-Alpha, Thermo Fisher) and Nano-FTIR (neaSNOM, aspect) were performed to determine the composition of the GQD and confirm the formation of an interface between GQD and hBN. Low voltage Cs aberration-corrected transmission electron microscopy (Titan Cube G2 60-300, FEI), operated at 80 kV with a monochromated electron beam, was used for electron energy loss spectroscopy (EELS) analysis. The spatial and energy resolutions for the EELS measurement are 2 nm and 1.5 eV, respectively.

**Device fabrication**

Optical images of the flakes at different stages of the fabrication process are presented on Supplementary Figure 14. The electrical response of the GQDs embedded in monolayer hBN was investigated by assembling vertical tunnel van der Waals heterostructures consisting of the stack of 30nm_hBN/Gr/2hBN/GQD-hBN/2hBN/Gr/20nm_hBN placed on an oxidised silicon wafer (300nm of $SiO_2$). Here, the bottom and top layers of hBN were used for the purpose of encapsulation. As the GQDs are embedded in the large area monolayer hBN on $Si/SiO_2$, the vertical heterostructure was assembled in two halves by adopting a mix of dry and wet flake transfer procedure: first, a stack of Si/ $SiO_2$/30nm_hBN/Gr/2hBN was prepared by standard flake exfoliation and dry transfer procedure. To prepare the other half, a stack of 2hBN/Gr/20nm_hBN was prepared by the dry pick up procedure using a PMMA membrane. This stack on the membrane was aligned and dropped on the GQD-hBN/$SiO_2$/Si substrate. To release the stack from Si/$SiO_2$, 8% PMMA was spun on the sample and Si/$SiO_2$ was etched using KOH solution. The floated membrane was thoroughly rinsed with DI water several times to remove KOH residues from the membrane. Finally, to complete the device, this membrane containing GQD-hBN/2hBN/Gr/20nm_hBN was dropped on the stack prepared in the first half.

For electrical characterisation of this vertical heterostructure, Cr/Au edge contacts were made on the top and bottom graphene layers using electron beam lithography followed by boron nitride etching, metal deposition, and lift-off process. Electrical characterisation of graphene contacts is presented in Supplementary Figure 15.

## Acknowledgements

This work was supported by the research fund (NRF-2017R1E1A1A01074493), the IBS (IBS-R-019-D1) and the grant (CASE-2013M3A6A5073173) from the centre for Advanced Soft Electronics under the Global Frontier Research Program through the National Research Foundation by the Ministry of Science and ICT, Korea., EU Graphene Flagship Program, European Research Council Synergy Grant Hetero2D, the Royal Society, Engineering and Physical Research Council (UK), US Army Research Office (W911NF-16-1-0279). S.O. acknowledges Ph.D. studentship from EPSRC funded Graphene NOWNANO CDT. J.Y. and A.M. acknowledges the support of EPSRC Early Career Fellowship EP/N007131/1.

# Supplementary Information

# Planar and van der Waals heterostructures for vertical tunnelling single electron transistors

*Gwangwoo Kim[1], Sung-Soo Kim[2], Jonghyuk Jeon[2], Seong In Yoon[1], Seokmo Hong[3], Young Jin Cho[4], Abhishek Misra[5,6], Servet Ozdemir[5], Jun Yin[5], Davit Ghazaryan[5,7], Mathew Holwill[5], Artem Mishchenko[5], Daria V. Andreeva[8], Yong-Jin Kim[9], Hu Young Jeong[10], A-Rang Jang[1,3], Hyun-Jong Chung[4], Andre K. Geim[5], Kostya S. Novoselov[5], Byeong-Hyeok Sohn[2] and Hyeon Suk Shin[1,3,9, 11]*

[1]Department of Energy Engineering, Ulsan National Institute of Science & Technology (UNIST), Ulsan 44919, Republic of Korea

[2]Department of Chemistry, Seoul National University, Seoul 08826, Republic of Korea

[3]Department of Chemistry, UNIST, Ulsan 44919, Republic of Korea

[4]Department of Physics, Konkuk University, Seoul 05029, Republic of Korea

[5]School of Physics and Astronomy, University of Manchester, Manchester M13 9PL, United Kingdom

[6]Department of Physics, Indian Institute of Technology Madras, Chennai, India

[7]Department of Physics, National Research University Higher School of Economics, Staraya Basmannaya 21/4, Moscow 105066, Russian Federation

[8]Department of Materials Science and Engineering, National University of Singapore, Singapore, 117575, Singapore

[9]Center for Multidimensional Carbon Materials, Institute of Basic Science (IBS), Ulsan 44919, Republic of Korea

[10]UNIST Central Research Facilities (UCRF), UNIST, Ulsan 44919, Republic of Korea

[11]Low Dimensional Carbon Material Center, UNIST, Ulsan 44919, Republic of Korea

[*]Present address: Carbon Composite Materials Research Center, Korea Institute of Science and Technology (KIST), Wanju 55324, Republic of Korea.

Correspondence and requests for materials should be addressed to H. S. S. (email: shin@unist.ac.kr), B. H. S. (email: bhsohn@snu.ac.kr) and K. S. N. (email: kostya@manchester.ac.uk ).

# Supplementary Figures

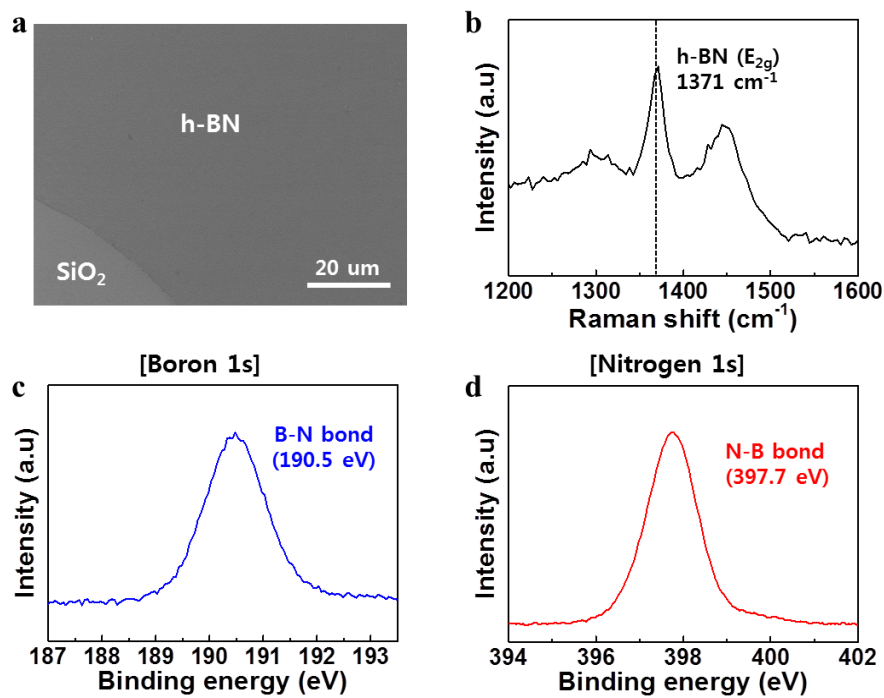

**Supplementary Figure 1. a, b,** The SEM image and Raman spectrum of the CVD grown pristine hBN monolayer. **c, d,** The corresponding XPS spectra: (c) Boron 1s, and (d) Nitrogen 1s.

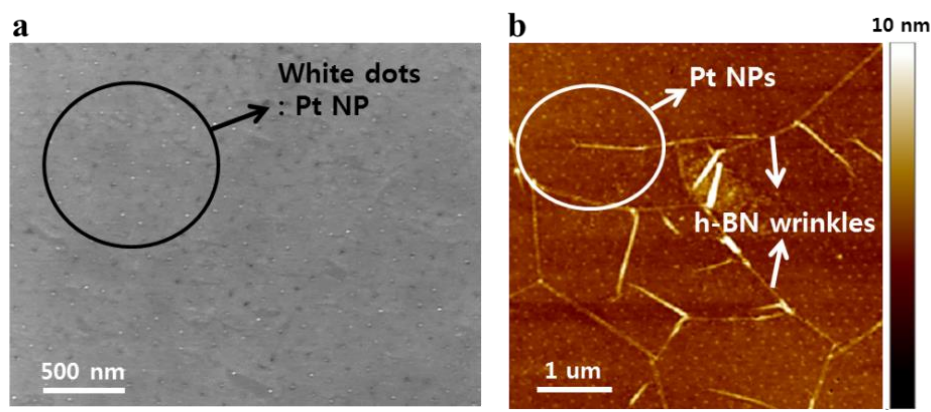

**Supplementary Figure 2. a, b,** The SEM and AFM images of the as-grown layer of GQD-hBN on an array of Pt NPs (7nm) spread over $SiO_2$ substrate.

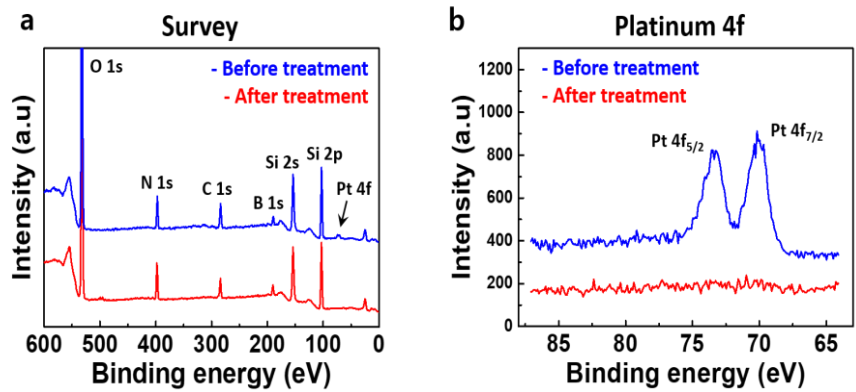

**Supplementary Figure 3. The XPS spectra of GQD-hBN planar heterostructure on SiO$_2$ substrate. a,** Survey, and **b,** Pt 4f spectra. Blue and red spectra are for as-prepared GQD/hBN on Pt NPs/SiO$_2$ substrate (before the aqua regia treatment) and the GQD-hBN after the aqua regia treatment to remove Pt NPs, respectively.

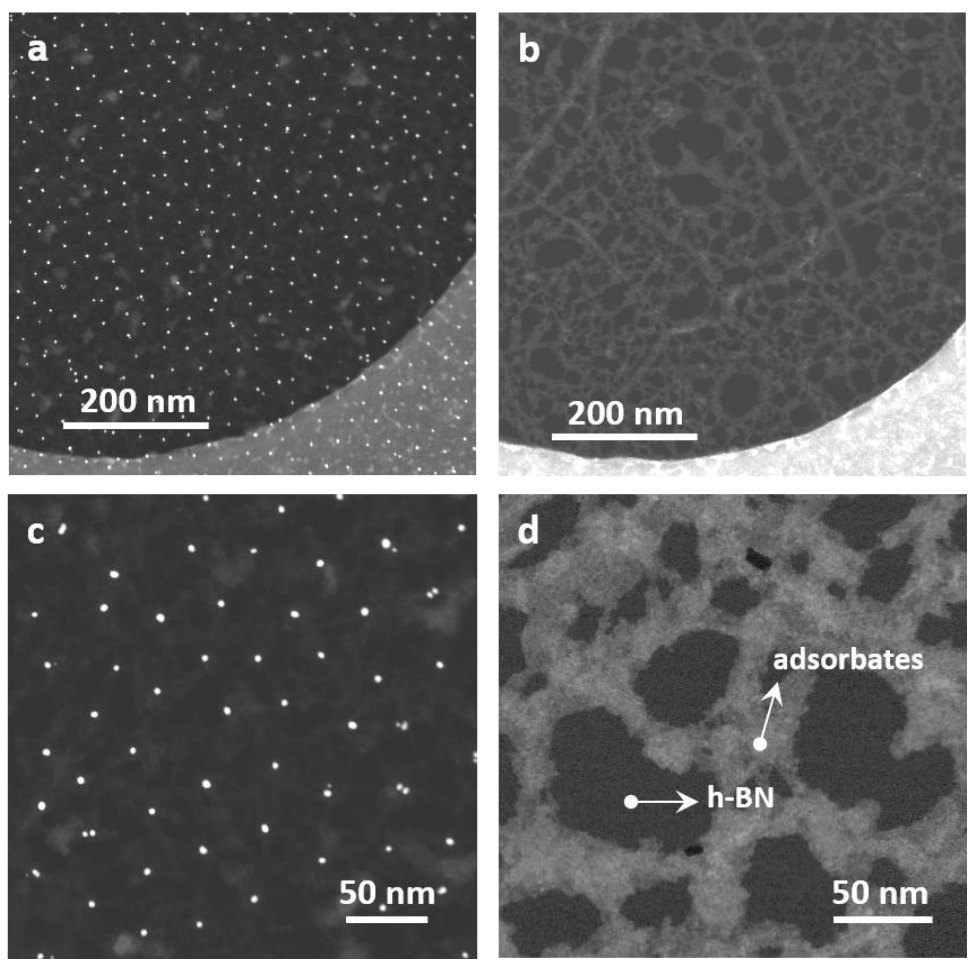

**Supplementary Figure 4. TEM images of GQD-hBN. a,c** Pt NPs in GQD-hBN. **b,d** The GQD-hBN after the aqua regia treatment. There are no Pt NPs after the treatment. The white dots in (a,c) are Pt NPs.

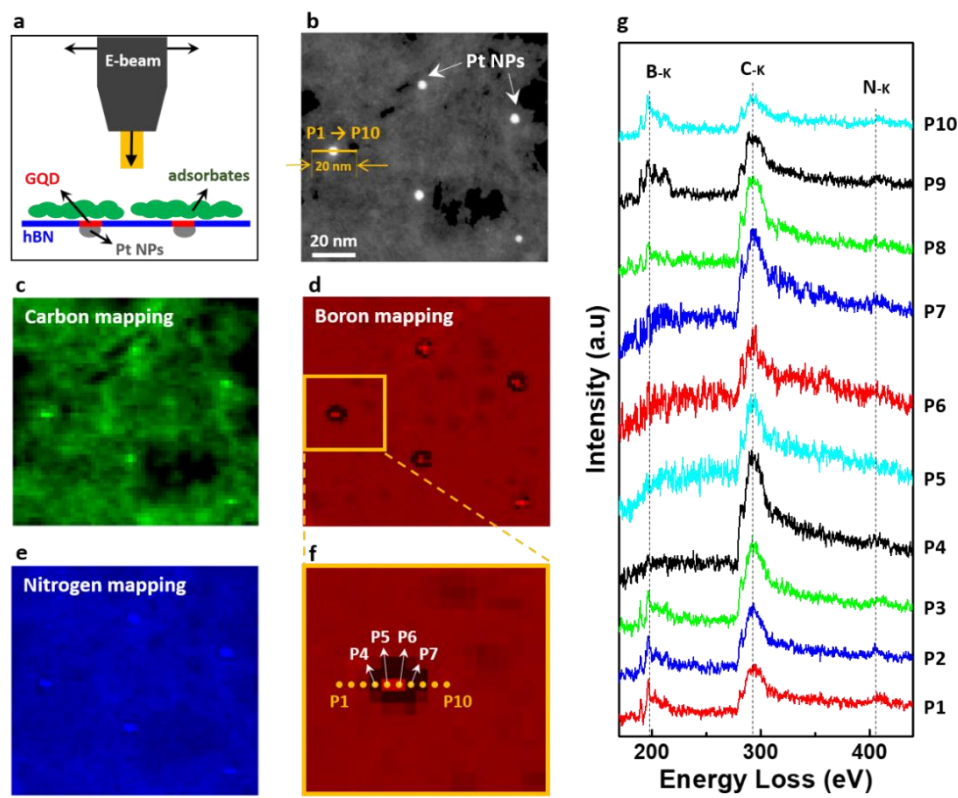

**Supplementary Figure 5. EELS spectrum of GQD-hBN. a** Schematics of EELS mapping of GQD-hBN on Pt NPs/SiO$_2$. **b**, TEM image of GQD-hBN on Pt NPs. The white dots are 7 nm Pt NPs. **c-e**, Corresponding EELS mapping images of (**c**) carbon, (**d**) boron, and (**e**) nitrogen, respectively. **f**, A magnified image marked in **d**. The boron signal was not detected at P4 to P7 where the Pt NP exist. Note that the points P5 and P6 with the strong signal are due to the strong background of Pt signal because we could not completely subtract the strong Pt background. Note that boron signal is absent in P5 and P6 (see panel **g**). In the nitrogen mapping image, the N signal is too low to be detected (see *Nano Lett.* 2013, *13*, 1834). **g**, The EELS spectra were obtained at different positions (yellow line, P1 to P10) with 2 nm spatial resolution in **f** by subtracting the background of the Pt signal from the original EELS spectra. The peak for Boron is not detected in P4, P5, P6, and P7, indicating conversion of BN to graphene. Note that GQDs in **c** and **g** are not distinguishable from carbon signal of many adsorbates.

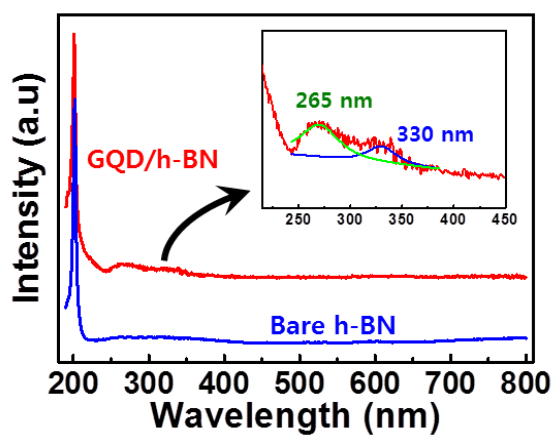

**Supplementary Figure 6.** The UV-vis absorbance spectra of pristine hBN and GQD-hBN planar heterostructure on quartz substrates. It shows the typical absorption band of hBN[1] located at 200 nm, and those of 265 nm and 330 nm for GQDs[2].

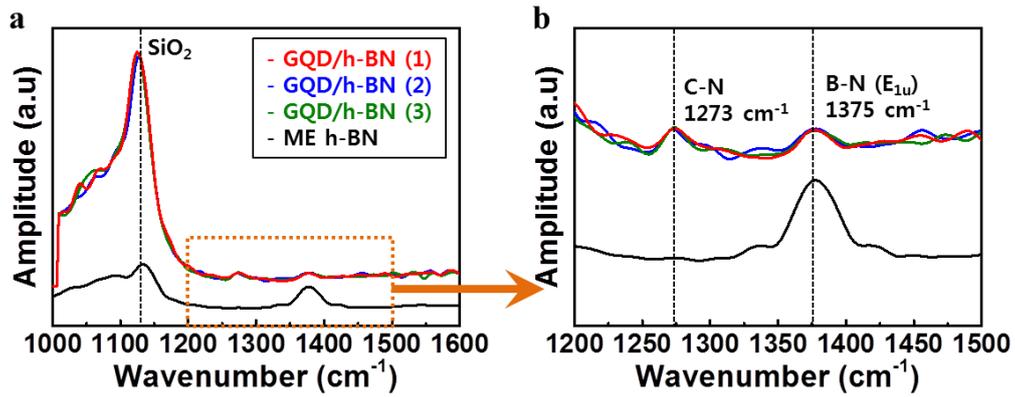

**Supplementary Figure 7. a,** The IR spectra for the GQD-hBN and mechanically exfoliated (ME) hBN (2nm thickness) on SiO$_2$ by means of AFM-IR. Three spectra at different points were measured for a GQD-hBN sample (1-3). **b,** Magnified spectra in the range of 1200-1500 cm$^{-1}$.

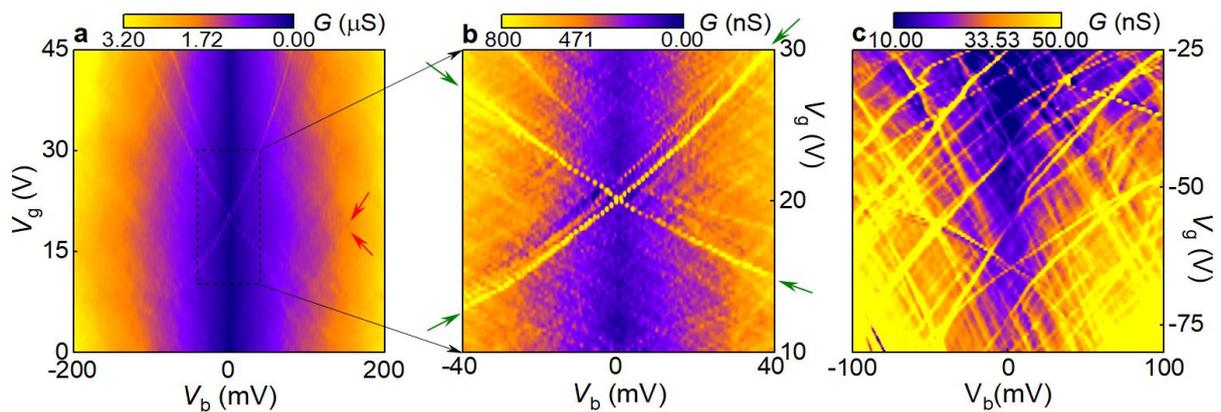

**Supplementary Figure 8**. **The hBN/Gr/2hBN/GQD-hBN/2hBN/Gr/hBN multi-channel single electron tunnelling transistors based on GQDs embedded in the hBN matrix. a, C**onductance $G(V_g, V_b)$ for the device with the GQDs of 7nm size prepared by the technique of self-assembly (measured at $T$=250mK). Red arrows indicate the edges of the Coulomb diamonds at the bias voltage of ≈160mV. **b,** The low excitation measurements of low bias region of (a), indicating the tunnelling events through the localised density of states in the middle hBN layer. The olive arrows denote the localised states with the energy 140meV below the Dirac point. **c,** Low-density non-periodic array of GQDs of 10nm size embedded in hBN layer placed in-between of hBN/Gr/2hBN/ and /2hBN/Gr/hBN heterostructures. T=250mK conductance $G(V_g, V_b)$ for a device with such GQDs and low quality exfoliated encapsulating bilayers of hBN. Because of the technology we used, there are roughly 4 times more 7nm (Supplementary Figure 10 (a)) GQD per unit area than there are 13nm (Figure 4 (b)) GQD, thus, there are more overlapping diamonds for 7nm GQD.

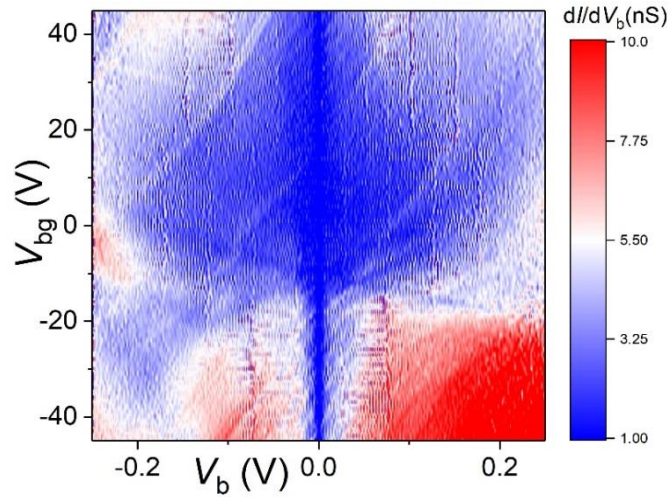

**Supplementary Figure 9**. *T*=250mK tunnelling conductance $G(V_b,V_g)$ of Si/SiO$_2$ substrate supported hBN/Gr/2hBN/CVD-hBN/2hBN/Gr/hBN heterostructure. The area of the device is 82μm$^2$. Note, the middle hBN monolayer was grown by CVD, but no GQD was formed on it. Note, significantly lower conductivity (even though the area of the device is significantly larger than for those presented in the main text) due to the absence of the additional conductance channels due to GQD. There is a small number of the impurity states, however, which might be originating from either defect in the CVD hBN, or due to contamination between the layers introduced during the fabrication.

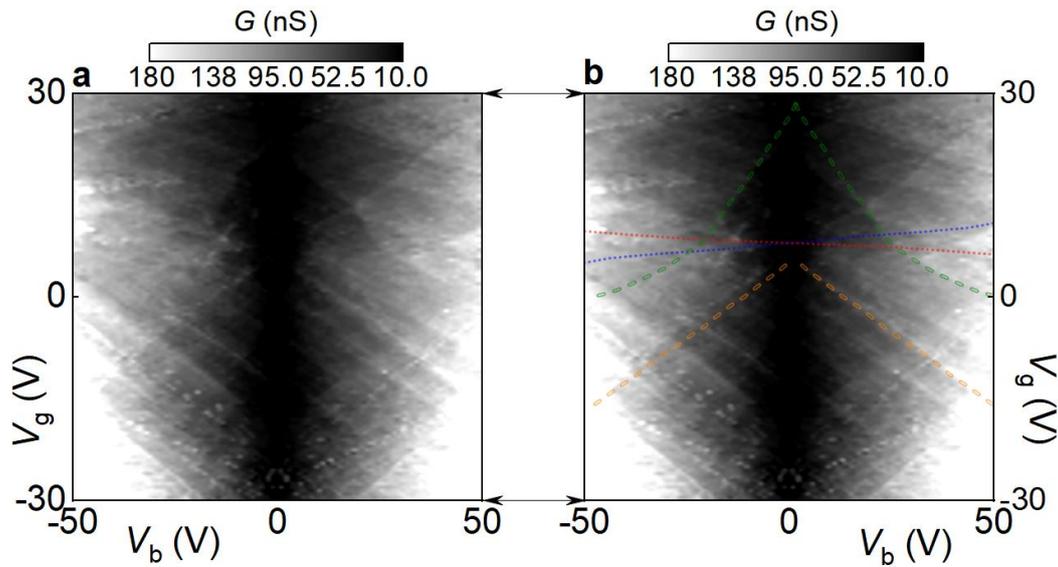

**Supplementary Figure 10. The peculiar shape of the Coulomb diamonds at low bias and gate voltages. a**, *T*=250mK low excitation $G(V_b, V_g)$ for the device with 7nm GQDs embedded in the central layer of CVD hBN. **b,** same as (a), but denoting a peculiar constitution of the lines forming Coulomb diamonds when the Fermi levels in either graphene contacts align with the Dirac points (green dashed lines). On the contrary, Orange dashed lines indicate the usual shape of the constituent lines of Coulomb diamonds when the Fermi levels are away from the DPs of both graphene layers, blue and red dotted lines.

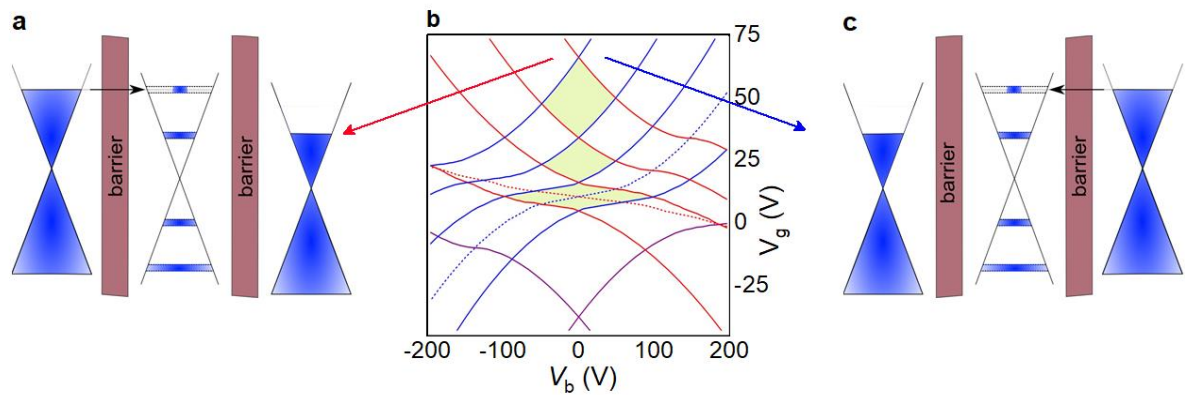

**Supplementary Figure 11. Single electron charging effect model. a-c,** Schematic representation of a single electron charging effect denoting particular approximations used in the modelling of electrostatic parameters of the resulting final heterostructure. **b,** same as Figure 6d of the main text.

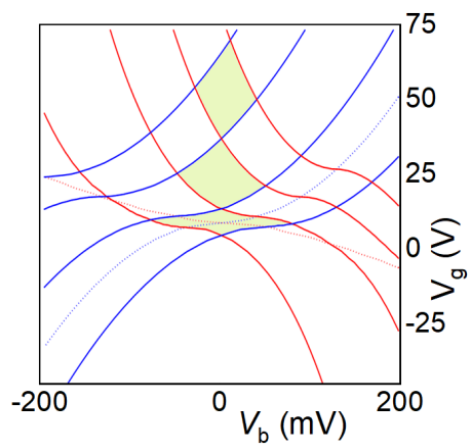

**Supplementary Figure 12. Single electron charging effect model.** Same as Figure 12b, except the chemical potential of the GQD-hBN layer is aligned with that in the source (bottom) electrode.

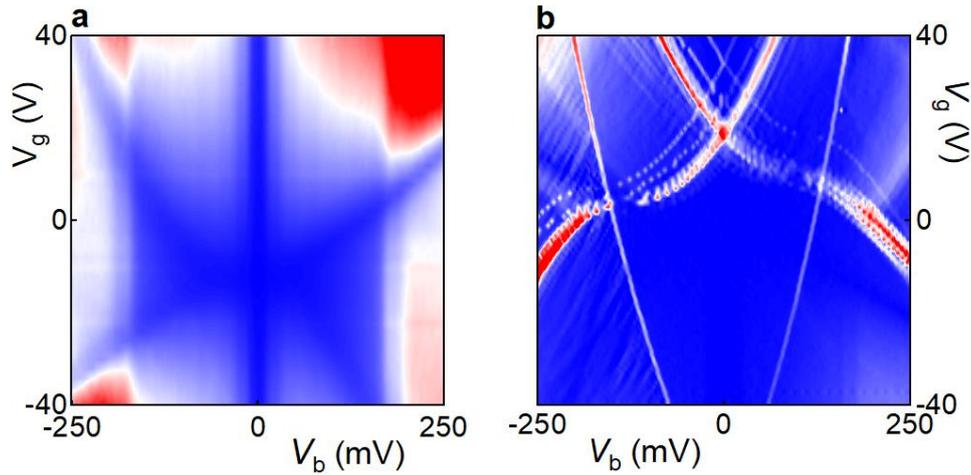

**Supplementary Figure 13.** *T*=1.5K tunnelling conductance $G(V_b, V_g)$ of Si/SiO$_2$ substrate supported Gr/hBN/Gr heterostructures. **a,** Tunnelling through pristine hBN trilayer mounted in-between two graphene monolayers (colour scale is blue to white to red, 20nS to 2μS to 4μS). Dark Blue *X* shaped region corresponds to the event of the passage of chemical potential through DPs of graphene layers; vertical features represent phonon-assisted resonant tunnelling process[3]. **b,** Tunnelling through impurity states of low-quality tetralayer hBN mounted in-between monolayer graphene electrodes (colour scale is blue to white to red, 0nS to 20nS to 40nS). Three various peaks in conductance (red and white) correspond to the tunnelling through three different localised states and all follow the square root dependence[4]. Tunnelling conductance of the heterostructures of hBN/Gr/hBN/Gr/hBN without the additional layer of the CVD hexagonal boron nitride with embedded GQDs is shown in the Supplementary Figure 11. Here, in the case of pristine high-quality hBN spacer mounted in between graphene layers, resonant tunnelling features (straight vertical lines at fixed voltages of $V_b$) occur through the assistance of both acoustic and optical phonons[3]. Notably, apart from the X shaped low conductivity region, which corresponds to the event of the passage of the chemical potential through the Dirac points of graphene contacts, there are no other features observed. On the other hand, in Supplementary Figure 10 (b), it was demonstrated that if a low-quality hBN is used as a tunnelling barrier in similar heterostructure - the tunnelling is dominated by the impurity states that are located in the middle of the barrier.

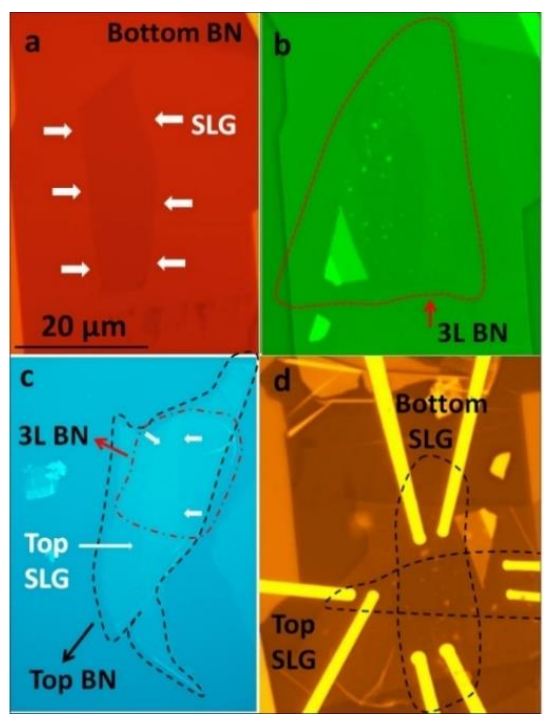

**Supplementary Figure 14. Fabrication procedure for van der Waals tunnel heterostructure comprising the stack of 20nm_hBN/Gr/3L_hBN/GQD-hBN/3L_hBN/Gr/10nm_hBN. a,** Single-layer graphene, indicated by white arrows, was transferred by flake peeling method on bottom hBN supported on Si/SiO$_2$ substrate. **b,** Trilayer hBN, outlined by red dashed line, was then transferred on graphene layer shown in (a). **c,** Separately a PMMA membrane was prepared with 10nm hBN, another graphene layer and trilayer hBN were picked up using this hBN. This PMMA membrane containing the stack of 3hBN/Gr/10_hBN was further aligned and dropped on GQD-hBN on Si/SiO$_2$. To release this stack from Si/SiO$_2$, wet transfer procedure following a standard KOH etching procedure was performed. **d,** Finally, the heterostructure of GQD-hBN/3hBN/Gr/10nm_hBN was aligned and transferred on 20nm_hBN /Gr/3hBN shown in (b). Contacts to the top and bottom graphene layers were made by standard electron beam lithography, as shown in (d). Scale bar shown in (a) is the same for all images.

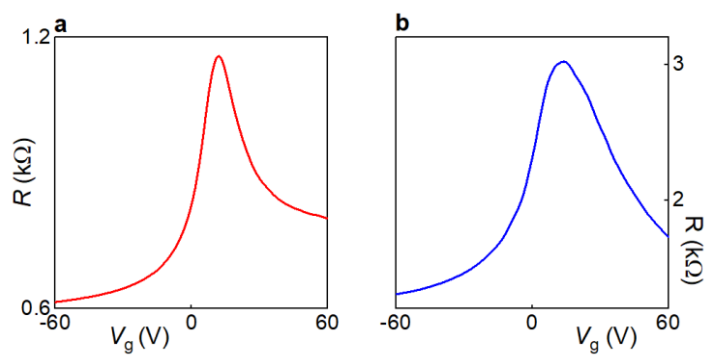

**Supplementary Figure 15**. **Resistance dependence of graphene on gate voltage for our van der Waals heterostructure with 13nm GQDs embedded in a central layer of hBN. a,** The 2-probe resistance measurements of the gate dependence of top monolayer graphene at $T$=0.25K. **b,** The 2-probe resistance measurements of the gate dependence of bottom monolayer graphene at $T$=0.25K.

# Supplementary Tables

| Size of GQD (nm) | $L_a$ (nm) |
|---|---|
| 7 nm | 8.98 nm |
| 10 nm | 10.12 nm |
| 13 nm | 11.79 nm |

**Supplementary Table 1. Graphitic domain size.** The integrated intensity ratio, $I_D/I_G$, was used to determine the in-plane crystallite size $L_a$ (nm) using the Tuinstra-Koenig relationship[5], as described in Supplementary Note 1 and Supplementary Equation (1).

| Dev. N | Fabrication technique of hBN-GQD layer | Cross-sectional area | Number of hBN layers used as spacers | GQD size extracted from SEM | Number of prominent localised electronic states, extracted from tunnelling spectroscopy | Estimated number of GQDs involved in the tunnelling measurements | The estimated charging energy for the GQDs (approximation) |
|---|---|---|---|---|---|---|---|
| 1 | Self-assembly | 60 µm² | 3hBN | 13nm | 10 | >80 | 80meV |
| 2 | Self-assembly | 30 µm² | 2hBN | 13nm | 3-4 | ~40 | 80meV |
| 3 | Self-assembly | 32 µm² | 2hBN | 10nm | 10 | ~40 | 100meV |
| 4 | Self-assembly | 40 µm² | 3hBN | 7nm | 5-6 | ~50 | 160meV |
| 5 | Pristine hBN | 71 µm² | 2hBN | N/A | 1-2 | N/A | 0meV |
| 6 | Pristine hBN | 80 µm² | 2hBN | N/A | N/A | N/A | 0meV |
| 7 | Non-periodic array | 6 µm² | 2hBN | 10nm | 10 | ~10 | 100meV |
| 8 | Non-periodic array | 33 µm² | 2hBN | 10nm | 6 | ~8 | 100meV |

**Supplementary Table 2. The functionality of the investigated devices.**

# Supplementary notes

## Supplementary Note 1. Estimation of the size of the GQD from Raman measurements

We used the Tuinstra-Koenig relationship[5] to estimate the size of the GQD ($L_a$) from our Raman measurements:

$$L_a(\text{nm}) = 2.4 \times 10^{-10} \, \lambda^4 \, (I_G/I_D) \qquad (1)$$

Here $\lambda$ is the wavelength of Raman excitation (532 nm), $I_G$ and $I_D$ – are the integrated intensities of the G and D peaks respectively.

## Supplementary Note 2. Modelling the electrostatic parameters of final heterostructures

The electrostatic equations for the vertical Gr/hBN/GQD-hBN/hBN/Gr heterostructures could be evaluated similar to[4], but with specific modifications. Here, Gr stands for the bottom and top electrodes of monolayer graphene, hBN stands for mechanically exfoliated layers of hexagonal boron nitride, and GQD-hBN stands for the CVD layer of hBN with an embedded GQDs.

Note, that the difference in modelling tunnelling though localised states and through GQD is that we do not allow charge accumulation on the localised state and we allow such charge accumulation on GQD due to finite capacitance.

At first, for the modelling of the tunnelling events through the localised edge states in the middle GQD layer we consider a model of a three-plate capacitor with the n-doped Si and two monolayers of graphene. Accounting the fact that the electric field generated by the n-Si gate electrode is partially screened by the bottom graphene layer, owing to the linear spectrum, we obtain

$$\begin{cases} eV_b = \mu_B - \mu_T - e^2 d_{SP} n_T(\mu_T)/\varepsilon_0 \varepsilon_{SP} \\ eV_g = \mu_B + e^2 d_{BG}(n_T(\mu_T) + n_B(\mu_B))/\varepsilon_0 \varepsilon_{BG} \end{cases} \qquad (2)$$

where $d_{SP}, d_{BG}, \varepsilon_{SP}, \varepsilon_{BG}$ are the thicknesses and dielectric constants of the spacer and back gate insulating films, respectively, and $\varepsilon_0$ is the vacuum permittivity. The relation of carrier density and chemical potential is given as $n_i(\mu_i) = \mu_i^2/\pi \hbar^2 v_F^2$, where $v_F$ is the Fermi velocity, and $\hbar$ is the Plank's constant. Considering the localised state as a level located at certain energy and a spatial position in the barrier developed by the gap of the spacer insulating film, we model resonant conditions of the tunnelling events through such states as

$$\mu_B = E_{LS} - (e^2 d_{SP} n_T(\mu_T)/\varepsilon_0 \varepsilon_{SP})(d_{LS}/d_{SP}), \qquad (3)$$
$$\mu_T = E_{LS} + (e^2 d_{SP} n_T(\mu_T)/\varepsilon_0 \varepsilon_{SP})(d_{SP} - d_{LS}/d_{SP}), \qquad (4)$$

where $E_{LS}$ is the energy of the specific level, localised state, (counted from the zero state of unpthe erturbed system - neutrality point of bottom layer electrode), and $d_{LS}$ is the spatial position of such a state in the barrier (counted from the bottom layer electrode).

Next, for the modelling of tunnelling events corresponding to the emergence of Coulomb diamonds due to the GQDs we consider the four-plate capacitor with n-Si, two monolayers of graphene, and middle GQD-hBN layer. Here, the GQD-hBN layer is introduced as an electrode with discrete energy levels (see Supplementary Figure 12), corresponding to the size quantization of GQDs. As indicated in the Supplementary Figure 12, there are two sets of lines; blue and red, corresponding to the different directions the single electron charging effect. To a good quantitative approximation, the chemical potential of the middle layer is considered to be aligned to the chemical potential of the top (drain) monolayer graphene electrode for both directions of the tunnelling (see Supplementary Figure 8a and red line in b). Such approximation allows for the analytical solution of the model. In such a scenario, accounting an additional screening arising from the middle GQD-hBN layer we obtain

$$\begin{cases} eV_b = \mu_B - \mu_M - e^2 d_{SP}(n_M(\mu_M) + n_T(\mu_T))/\varepsilon_0 \varepsilon_{SP} \\ eV_g = \mu_B + e^2 d_{BG}\left(n_M(\mu_M) + n_B(\mu_B)\right)/\varepsilon_0 \varepsilon_{BG} \\ \mu_M = \mu_T + e^2 d_{SP2}(n_T(\mu_T))/\varepsilon_0 \end{cases} \quad (5)$$

where $d_{SP2} = d_{SP}/2$, and $\mu_M$ is the chemical potential of the middle GQD-hBN layer. In this case, the resonant tunnelling condition for the tunnelling direction bottom to top are given as

$$\mu_B = (e^2 d_{SP2}(n_M(\mu_M) + n_T(\mu_T)/\varepsilon_0 \varepsilon_{SP}) + E_i, \quad (6)$$

where $E_i$ are the discrete energy levels in the middle GQD-hBN layer, and $d_{SP2}$ is the thickness of the hBN barrier between the bottom graphene and GQD-hBN layer. Likewise, to model the resonant conditions for an opposite direction of the tunnelling, one needs to consider the relation between the chemical potential of the top layer of graphene and the emerged electrostatic field to the energy levels $E_i$ of a middle GQD-hBN layer (see Supplementary Figure 12c, and blue lines in b).

To prove that our assumption of aligning the chemical potential of the GQD-hBN layer with that in the drain electrode does not introduce any significant qualitative errors, we modelled the opposite extreme situation, when the chemical potential of the GQD-hBN layer is aligned to that in the source (bottom) electrode. One can see from Supplementary Figure 13 that the position and the shape of the diamonds are very much the same as in the previous case. Thus, our approximate model, which could be solved analytically, provides a good qualitative and quantitative description of tunnelling through GQDs.

## Supplementary References